\pdfoutput=1 
\documentclass[10pt]{article}
\usepackage[utf8]{inputenc}
\usepackage[T1]{fontenc}
\usepackage[top=1in,bottom=1in,left=1in,right=1in]{geometry}
\usepackage{amsmath,amsfonts}
\usepackage{amssymb}
\usepackage{stmaryrd}
\usepackage{graphicx}
\usepackage{url}
\usepackage{soul}
\usepackage{multirow}
\usepackage{hyperref}
\usepackage{theorem}

\usepackage[lined]{algorithm2e}

\def\S{\mathcal{S}}
\def\F{\mathcal{F}}

\begin{document}
	\title{Empirical Voronoi Wavelets} 
	\author{J\'er\^ome Gilles\footnote{Department of Mathematics \& Statistics, San Diego State University, 5500 Campanile Dr, San Diego, CA 92182, USA, jgilles@sdsu.edu}}

\maketitle
\pagestyle{myheadings}
\markboth{J\'er\^ome Gilles}{Empirical Voronoi Wavelets}

	\begin{abstract}
		
		Recently, the construction of 2D empirical wavelets based on partitioning the Fourier domain with the watershed transform has been proposed. If such approach can build partitions of completely arbitrary shapes, for some applications, it is desirable to keep a certain level of regularity in the geometry of the obtained partitions. In this paper, we propose to build such partition using Voronoi diagrams. This solution allows us to keep a high level of adaptability while guaranteeing a minimum level of geometric regularity in the detected partition.
		\vspace{5pt}
	\end{abstract}

	\section{Introduction}\label{sec:intro}
Empirical wavelets have been proposed in \cite{EWT1D} in the 1D case, and then extended to 2D in \cite{EWT2D} as an alternative to the empirical mode decomposition \cite{Huang1998}. Its purpose is to build data-driven wavelets, i.e. a family of wavelets which is designed based on the content of the original signal/image to analyze. The corresponding wavelet filter bank aims at extracting the harmonic modes (i.e. amplitude modulated - frequency modulated components) plus some residue. This is achieved by considering that the expected modes should have a compact support (or at least are rapidly decreasing outside a compact support) in the Fourier domain. Therefore, the adaptability is obtained by detecting the supports of each mode instead of following some prescribed rule like classic wavelets. A wavelet filter is build for each support providing us the sought wavelet filter bank. A theoretical analysis of such construction in the 1D case is available in \cite{cewt}, considering arbitrary partitioning of the Fourier domain. If in 1D, partitions are made of intervals, partitions in 2D can have more variability in terms of their geometry. For instance, in \cite{EWT2D}, several types of geometries have been considered like rectangular boxes (analogous to a tensor approach), concentric rings centered at the origin (to represent Littlewood-Paley type operators), and polar wedges (to mimic the behavior of curvelets). A higher degree of flexibility have been achieved in \cite{ewwt}, where partitions of arbitrary shapes are detected thanks to a watershed transform. Such level of adaptability is desirable for many applications, however it can lead to non-smooth geometries, affecting the degree of regularity of the wavelets themselves, which is frequently a desirable property for particular analyses. In this paper, we propose an alternative type of partitions based on Voronoi diagrams. This solution provides a trade-off between a high level of adaptability while keeping some simple geometric constraint on the partition to keep good properties of the obtained wavelets.

The remaining of the paper is organized as follows. Section~\ref{sec:ew} gives a brief reminder about empirical wavelets, in particular the 2D case. Section~\ref{sec:evw} describes the Voronoi based empirical wavelets. Some experiments will be presented in Section~\ref{sec:exp} while conclusions will be given in Section~\ref{sec:conc}.

\section{Empirical wavelets}
\label{sec:ew}
Empirical wavelets have proven to be very efficient in different problems from science and engineering, see for instance  \cite{microscopy,supewttexture,ewttexture,ewtapp1,ewtapp3} to cite only a few. Their construction is originally inspired by the Empirical Mode Decomposition \cite{Huang1998}. It aims at writing a signal $f$ as the sum of harmonic modes $f_k$ (i.e. amplitude modulated-frequency modulated components) and some residue $r$:
$$f(x)=r(x)+\sum_{k=1}^N f_k(x).$$
The key assumption is that the Fourier transform, $\hat{f}_k$, of each mode has a compact support, or is at least rapidly decaying outside of a compact support. Each wavelet filter, $\psi_k$, is then built on top of each support. The expected modes are obtained by filtering the signal $f$ by $\psi_k$, i.e.
$\hat{f_k}(\xi)=\hat{f}(\xi)\hat{\psi}_k(\xi)$, where $\xi$ is the frequency, and then the inverse Fourier transform is applied to get $f_k$. Note that we will denote $\psi_0$ the scaling function that extracts the residue $r$ (i.e. $f_0=r$). Empirical wavelets follow the same principle as classic wavelet except that the supports of each wavelet filter in the Fourier domain are not given by a given rule (like the dyadic decomposition) but are detected from the spectrum of $f$ itself. In practice, given a function $f$, we compute its magnitude spectrum, $|\hat{f}|$, then partition the Fourier domain to obtain the supports of the expected harmonic modes. Equipped with this partition, we build the wavelet filters and finally decompose $f$. This process describes the empirical wavelet transform. Note that this transform is not linear since the support detection step is, in general, not linear.

\begin{figure}[t]
\begin{tabular}{cccc}
\includegraphics[width=0.23\textwidth]{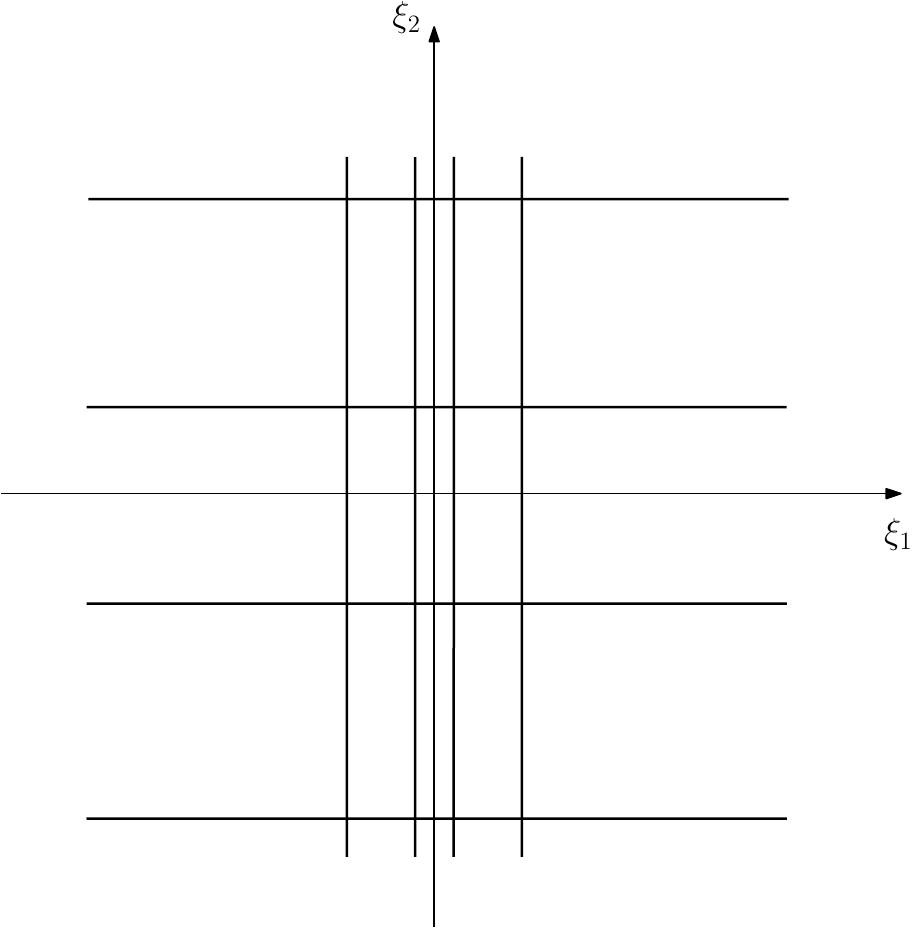} &
 \includegraphics[width=0.23\textwidth]{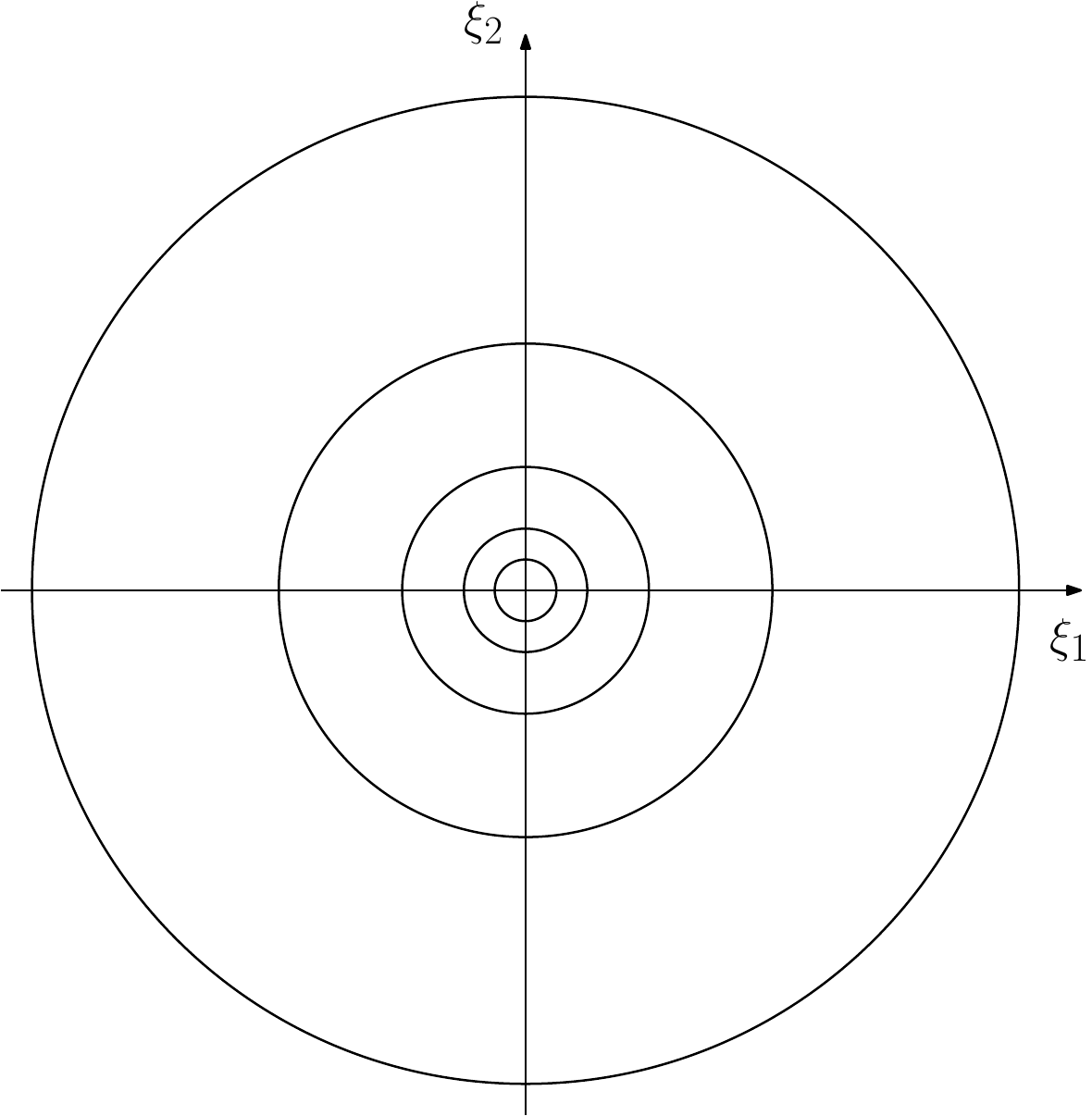} &
 \includegraphics[width=0.23\textwidth]{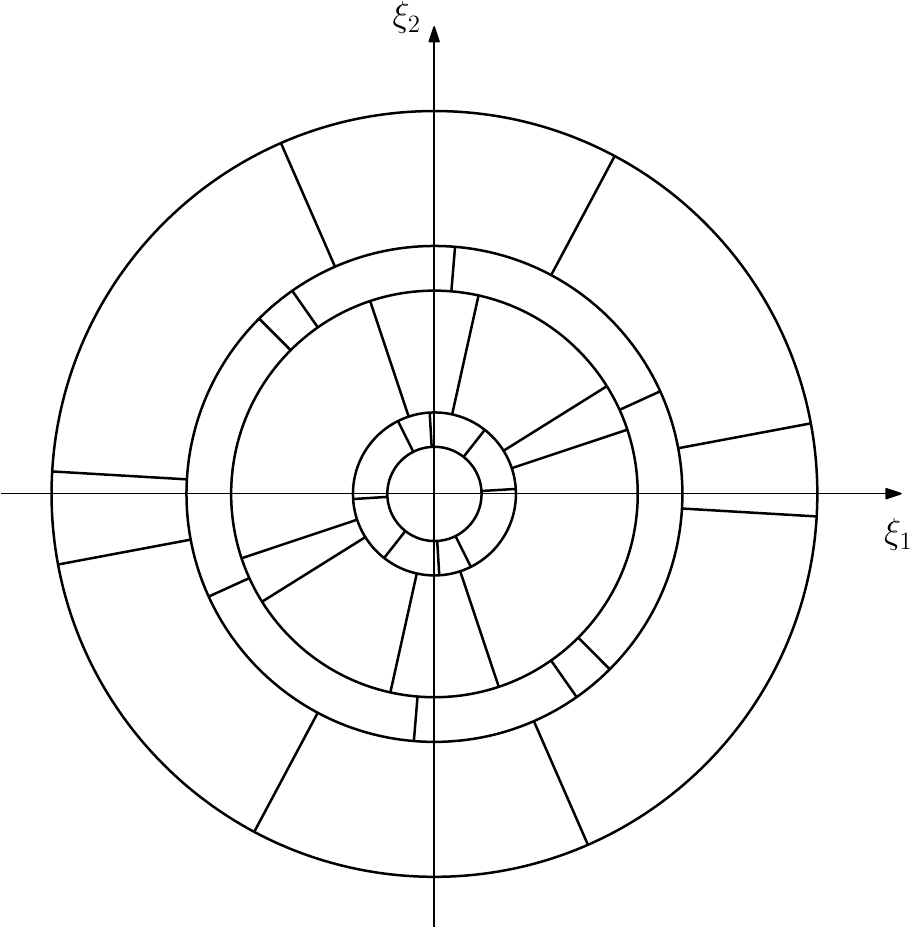} &
 \includegraphics[width=0.23\textwidth]{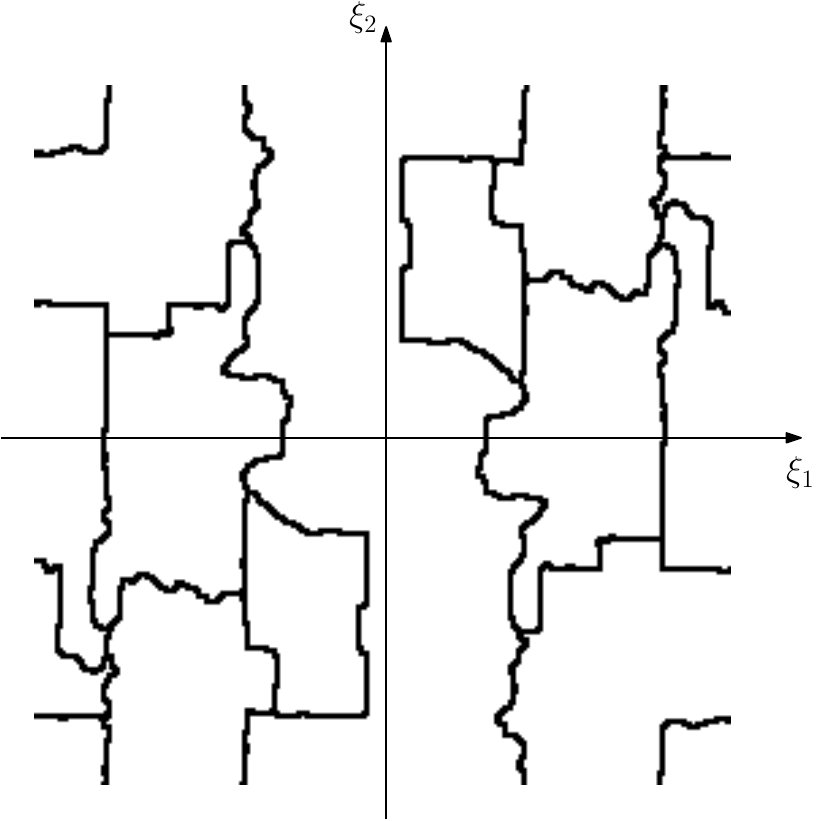} \\
 a) & b) & c) & d)
\end{tabular}
\caption{Existing 2D partitions of the Fourier domain. These different types of partitions correspond to a) tensor wavelets, b) Littlewood-Paley wavelets, c) curvelet type, d) watershed wavelets.}
\label{fig:part}
\end{figure}

If in 1D, partitions of the Fourier domain are collections of intervals, in 2D, the partition cells can have very different geometries. For instance, in \cite{EWT2D}, the authors have re-visited some existing constructions of classic wavelets, and have shown that building empirical versions of them is equivalent to partition the 2D Fourier domain with 1) rectangular boxes those edges are parallel to the frequency axis, 2) concentric rings centered around the origin, 3) polar wedges. Each of these types of partitions correspond respectively to tensor wavelets (Figure~\ref{fig:part}.a), Littlewood-Paley wavelets (Figure~\ref{fig:part}.b), and curvelets (Figure~\ref{fig:part}.c). These partitions have strong geometric constraints since they are based on boxes, rings and angular sectors. A higher level of adaptability has been reached in \cite{ewwt} where the authors proposed to use a watershed transform \cite{watershed1,watershed2,watershed3} to find the lowest level lines in the Fourier domain that separate the expected supports, see Figure~\ref{fig:part}.d. Such approach removes all geometric constraints on the shape of the partition cells. However, if such flexibility is desirable for some applications, it also has some drawbacks. More specifically, the curves corresponding to the cell edges may lack some smoothness which will directly impact the level of regularity of the built wavelets, which can be an issue in particular circumstances. To mitigate such issue, we propose in the next section to use Voronoi partitions. This solution allows us to keep a comparable level of flexibility, since we use the same seeds than in the watershed case to find the Voronoi cells; and the partition geometry is smoother since the cells edges are made of linear segments.

\section{Empirical Voronoid wavelets}
\label{sec:evw}% \renewcommand{\rmdefault}{phv} % Arial

In this section, we give the details on the construction of Empirical Voronoi Wavelets (EVW). In a nutshell, the different steps are: 1) detect the position of meaningful harmonic modes within the magnitude spectrum, 2) create the Voronoi partition, 3) build the wavelet filters accordingly to each Voronoi cell. Finally, the transform is obtained by performing each individual filtering. Hereafter, we provide details about these different steps.

\subsection{Detection of harmonic mode positions}
\begin{figure}[t]
\centering
\begin{tabular}{ccccc}
\includegraphics[scale=0.3]{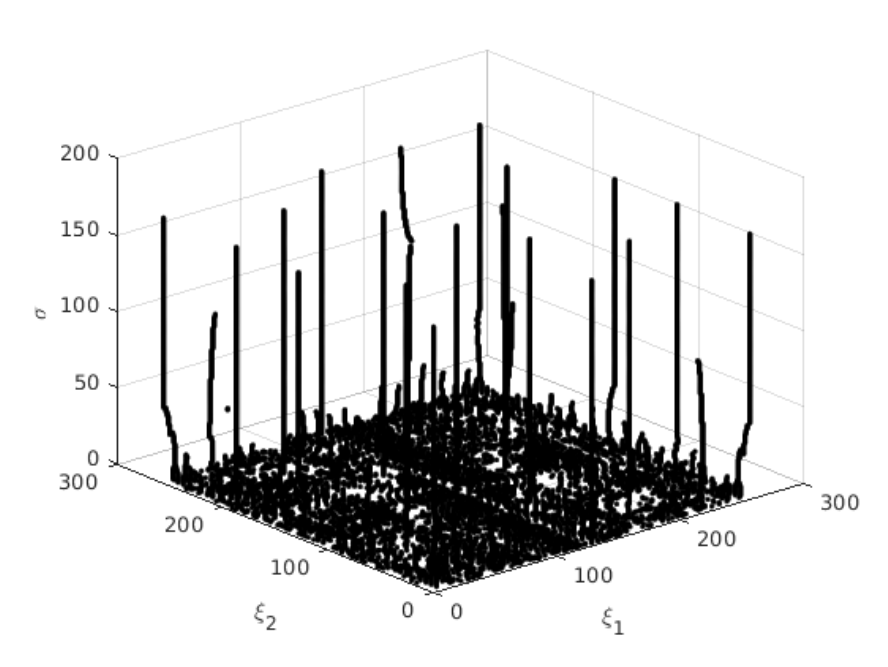} &
\hspace{3mm} &
\boxed{
    \includegraphics[scale=0.3]{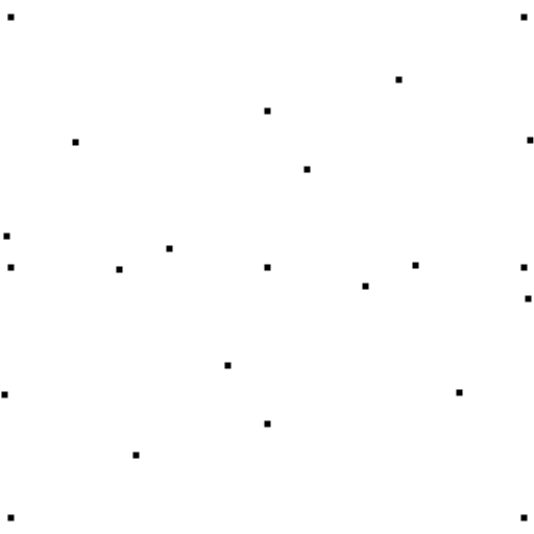}} &
\hspace{3mm} &
\boxed{
    \includegraphics[scale=0.3]{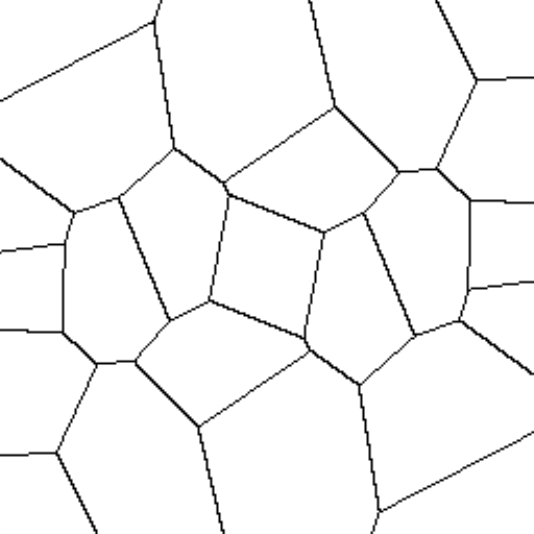}}
\\

a) & & b) & & c)
\end{tabular}
\caption{a) existence of local maxima in the scale-space representation. Each curve correspond to one originally detected maxima. The vertical axis corresponds to the scale parameter $\sigma$. b) positions of maxima $\{\xi_n^0\}_{n\in\Lambda}$ corresponding to the meaningful modes. c) the Voronoi partition associated with $\{\xi_n^0\}_{n\in\Lambda}$.}
\label{fig:ss}
\end{figure}

To detect the position of the meaningful harmonic modes within the magnitude spectrum, $|\hat{f}|$, we use the same method as in \cite{ewwt}. It consists in building a scale-space representation of $|\hat{f}|$:
$\S(\xi,\sigma)=(|\hat{f}|\ast g_\sigma)(\xi)$,
where $g_\sigma$ is a Gaussian kernel with variance $\sigma$. For each value of $\sigma$ (we take the convention that $g_0=\delta$, the Dirac function), we detect the set of local maxima, denoted $\{\xi_n^\sigma\}_{n=1}^{N_\sigma}$, in $\S$, where $N_\sigma$ is the number of such local maxima for $\sigma$. By increasing $\sigma$, the spectrum becomes smoother removing the small variations within it, hence $N_\sigma$ is decreasing (this property comes from one axiom of the scale-space theory that states that no extrema can appear while $\sigma$ increases). We can then obtain a binary scale-space representation that is zero everywhere except where local maxima were detected, see Figure~\ref{fig:ss}.a. The main idea is to notice that the maxima that do correspond to the expected meaningful modes are the ones corresponding to the ``longest'' curves in that representation. Therefore, if we denote 
$l_n=\arg\max_\sigma \{\xi_n^\sigma\;\text{exist}\}$ the length of the curve associated to $\xi_n^0$, we can compute the histogram of $\{l_n\}_{n=1}^{N_0}$. This histogram will be bimodal: one mode will mostly count for the shortest curves while the second one for the longest ones. Then, we use Otsu's algorithm \cite{Otsu1979} to automatically find a threshold, $T$, that separates these two modes. The indices of the sought longest curves are then given by $\Lambda=\{n\, |\, l_n>T\}$, the position of the meaningful modes $\{\xi_n^0\}_{n\in\Lambda}$ extracted from Figure~\ref{fig:ss}.a are depicted in Figure~\ref{fig:ss}.b.

\subsection{Voronoi partitioning}
The next step consists in using the set $\{\xi_n^0\}_{n\in\Lambda}$, found previously, as the seeds of a Voronoi partitioning algorithm \cite{voro1}. Each position in the domain is tagged with the label of the closest maxima position (we used the Euclidean distance). Note that, in the numerical implementation, if a given position is at equal distance from two maxima, some rule must be implemented to preserve the central symmetry that is expected when real images are processed. 
The Voronoi partition corresponding to the set of meaningful maxima depicted in Figure~\ref{fig:ss}.b is given in Figure~\ref{fig:ss}.c. To enforce a real transform, we pair together the Voronoi cells that are symmetric with respect to the origin. 

\subsection{Empirical Voronoi Wavelet transform}
The construction of the wavelet filters follow the same procedure as in \cite{ewwt}. Given a Voronoi cell $\Omega$, if we denote $\partial\Omega$ its edge, we define a distance transform by
\begin{equation}
\label{eq:distanceTransform}
D_\Omega(k,l) =\begin{cases} 
\frac{2\pi}{\mathcal{N}}\min_{(p,q)\in \partial\Omega} \big(d(k,l,p,q)\big) \quad &\text{ if } (k,l) \in \Omega \\
-\frac{2\pi}{\mathcal{N}}\min_{(p,q)\in\partial\Omega} \big(d(k,l,p,q)\big) &\text{ if } (k,l) \notin \Omega
\end{cases},
\end{equation}
where $d(k,l,p,q)$ is the quasi-Euclidean distance: 
\begin{equation}
\label{eq: qedist}
d(k,l,p,q) = \begin{cases}
(\sqrt2 -1) |q-l| + |p-k| \qquad &\text{ if }\; |p-k| \geq |q - l| \\
(\sqrt2 -1) |p-k| + |q-l|  &\text{ if }\; |p - k|<|q-l|.
\end{cases}
\end{equation}
The corresponding empirical Voronoi wavelet filter, $\widehat\psi_\Omega$ is then defined in the Fourier domain by
\begin{equation}
\label{eq:2dempiricalwavelet}
\widehat\psi_\Omega(k,l) = \begin{cases} 
1  &\text{ if }  D_\Omega(k,l) > \tau \\ 
\cos\left( \frac{\pi}{2} \beta\left( \frac{\tau-  D_\Omega(k,l)}{2\tau} \right)\right) \quad &\text{ if } D_\Omega(k,l) \leq |\tau| \\ 
0 &\text{ if }   D_\Omega(k,l) < -\tau,
\end{cases}
\end{equation}
where $\tau$ defines the width of a transition area along $\partial\Omega$ and $\beta(x) = x^4(35-84x+70x^2-20x^3)$. The Empirical Voronoi Wavelet transform (EVWT) is summarized in Algorithm~\ref{alg:evwt} (we denote $\F$ and $\F^{-1}$ the Fourier transform and its inverse, respectively).

\begin{algorithm}[t]
    \DontPrintSemicolon
    \LinesNumbered
    
    \SetKwInOut{Input}{Input}
    \SetKwInOut{Output}{Output}
 \caption{Empirical Voronoi Wavelet Transform}
    \label{alg:evwt}
    
    \Input{image $f$}
    \Output{set of EVW filters $\{\widehat\psi_{\Omega_k}\}$, set of wavelet coefficients $\{f_k\}$}
    \BlankLine
    \emph{$\hat{f} \leftarrow \F(f)$}\;
    \emph{Detect position of harmonic modes $\{\xi_n^0\}_{n\in\Lambda}$ from $|\hat{f}|$}\;
    \emph{Create the Voronoi partition $\{\Omega_k\}_{k=1}^N$ from the seeds $\{\xi_n^0\}_{n\in\Lambda}$}\;
    \For{$k=1$ to $N$}{
        \emph{Build $\widehat\psi_{\Omega_k}$ using Eq.\eqref{eq:2dempiricalwavelet}}\;
        \emph{Extract wavelet coefficients $f_k=\F^{-1}(\hat{f}\;\widehat\psi_{\Omega_k})$}\;
    }
\end{algorithm}

It is straightforward to see that Proposition~1 in \cite{ewwt} remains valid in the present work since a Voronoi partition can be seen as a particular case of the more general partition considered in \cite{ewwt}. Therefore, the set $\{\psi_{\Omega_k}\}$ forms a frame. The direct consequence is the guaranty of the existence of the inverse transform by constructing the dual frame $\{\tilde{\psi}_{\Omega_k}\}$ via 
$$\widehat{\tilde\psi}_{\Omega_k}=\frac{\widehat\psi_{\Omega_k}}{\sum_{k=0}^N |\widehat\psi_{\Omega_k}|^2}.$$ 
The inverse transform is thus given by 
$$f=\F^{-1}\left(\sum_{k=0}^N \hat{f}_k\;\widehat{\tilde\psi}_{\Omega_k}\right).$$

\section{Experiments}
\label{sec:exp}
\begin{figure}[t]
\begin{tabular}{cccc}
    
&
\includegraphics[width=0.22\textwidth]{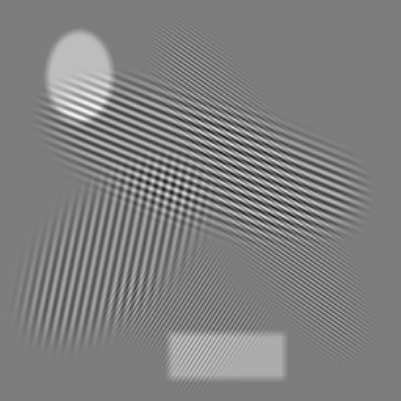} &
\includegraphics[width=0.22\textwidth]{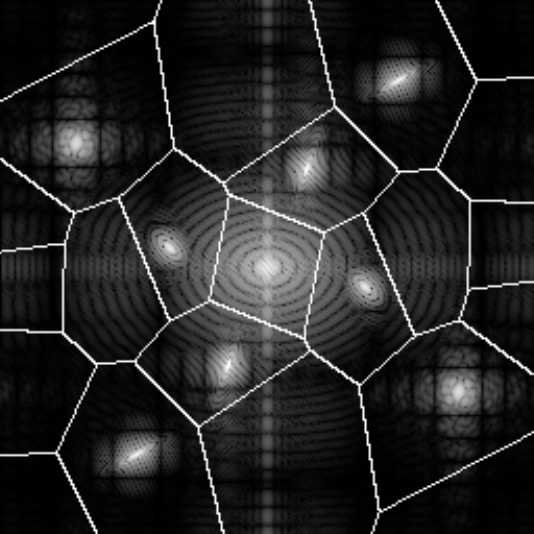} &
\\
    
\boxed{\includegraphics[width=0.22\textwidth]{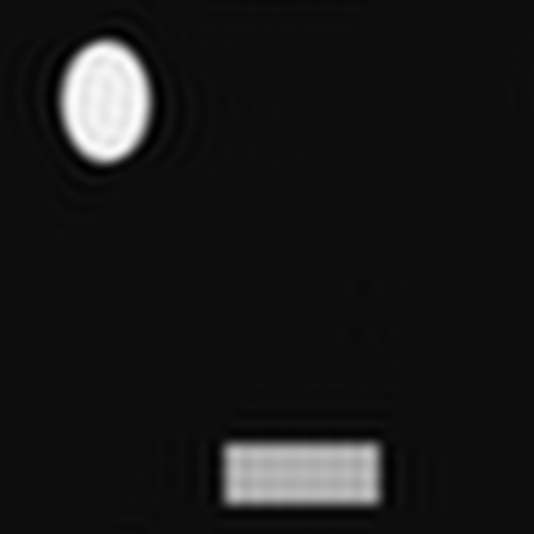}} &
\includegraphics[width=0.22\textwidth]{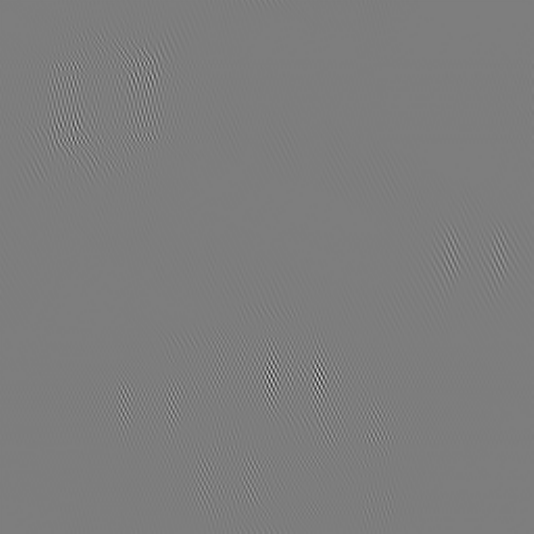} &
\includegraphics[width=0.22\textwidth]{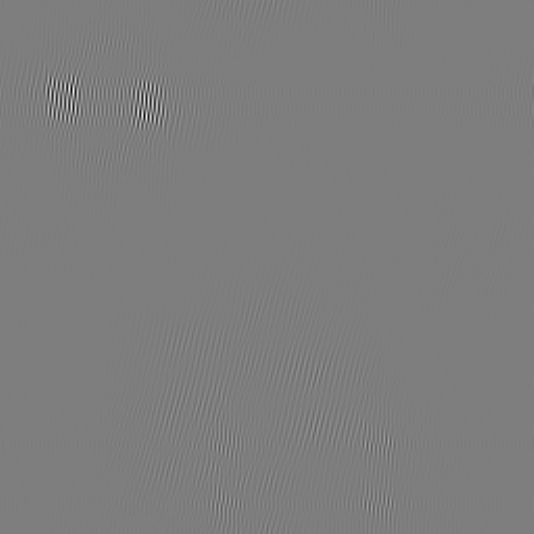} &
\includegraphics[width=0.22\textwidth]{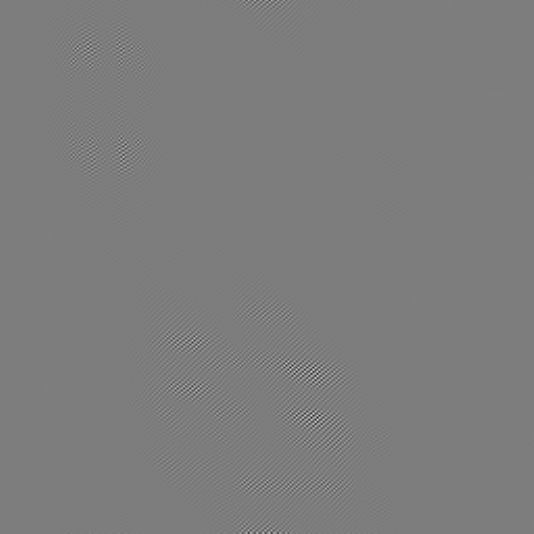} \\

\includegraphics[width=0.22\textwidth]{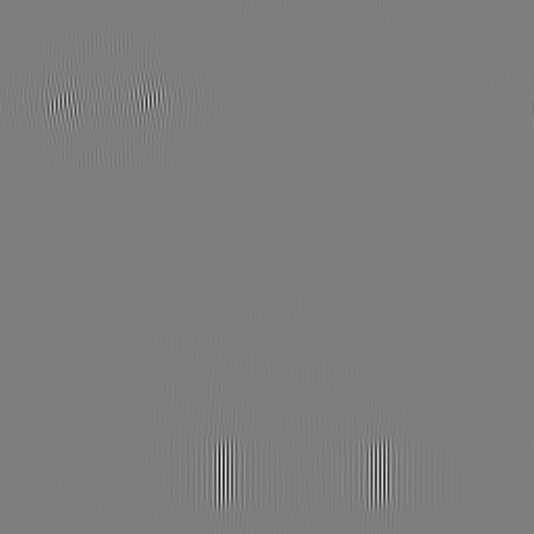} &
\includegraphics[width=0.22\textwidth]{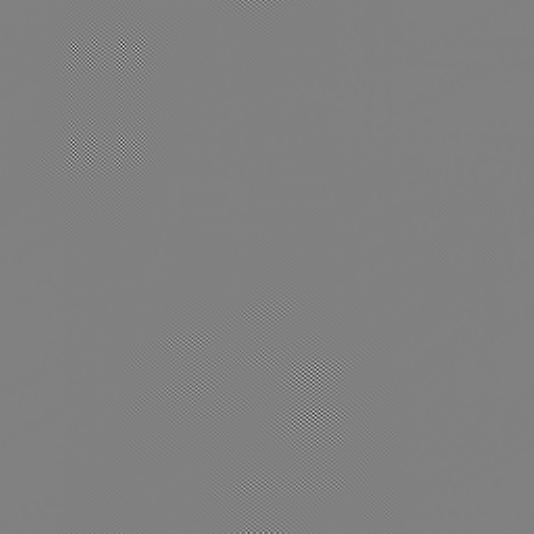} &
\boxed{\includegraphics[width=0.22\textwidth]{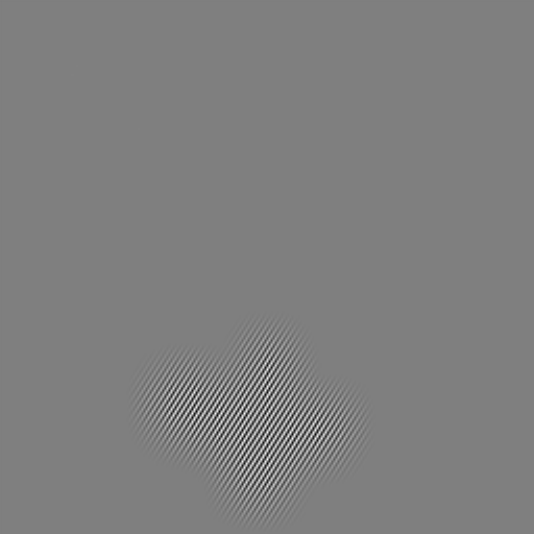}} &
\includegraphics[width=0.22\textwidth]{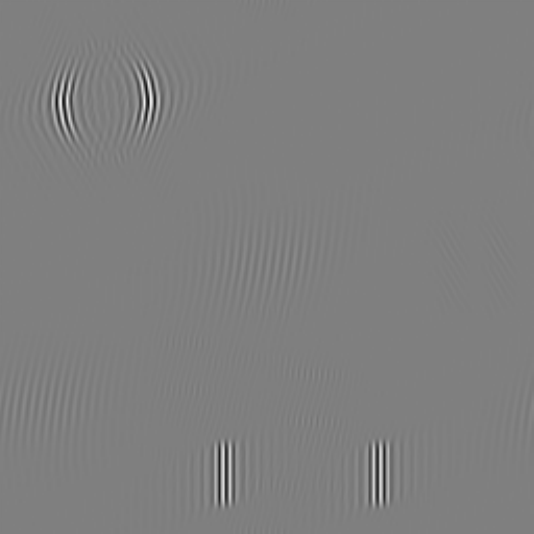} \\

\boxed{\includegraphics[width=0.22\textwidth]{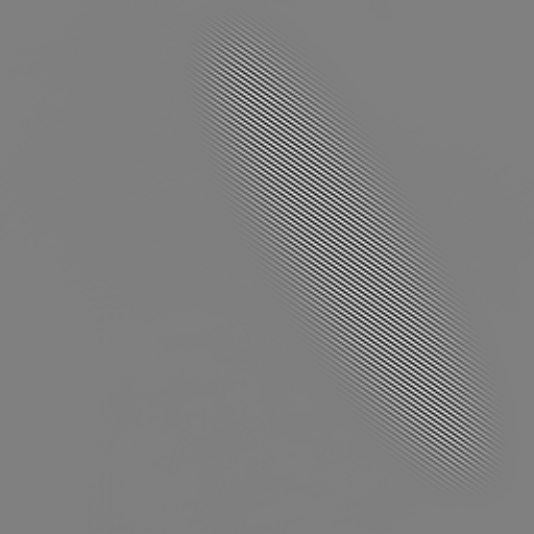}} &
\boxed{\includegraphics[width=0.22\textwidth]{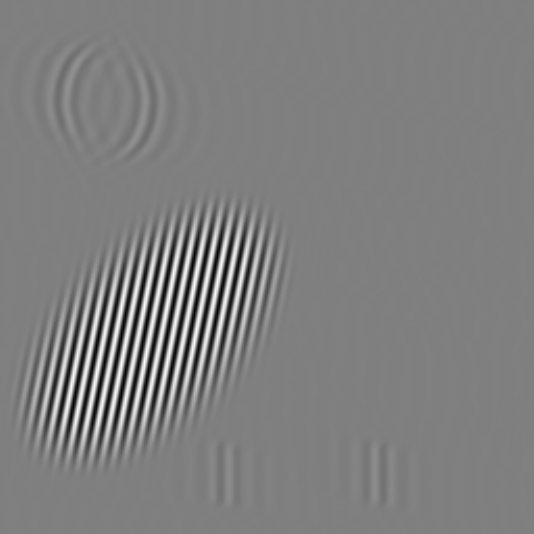}} &
\boxed{\includegraphics[width=0.22\textwidth]{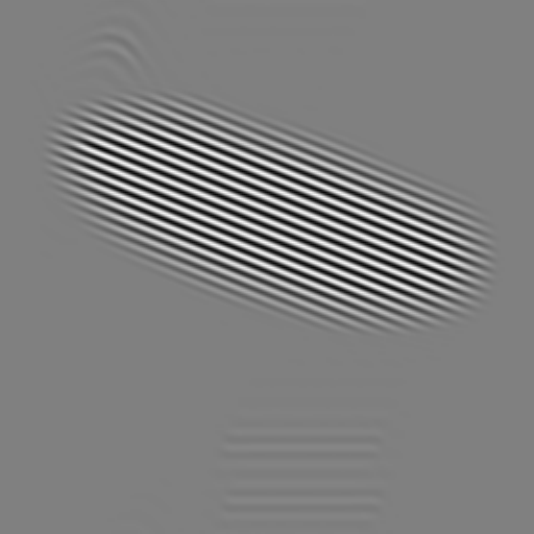}} &
\includegraphics[width=0.22\textwidth]{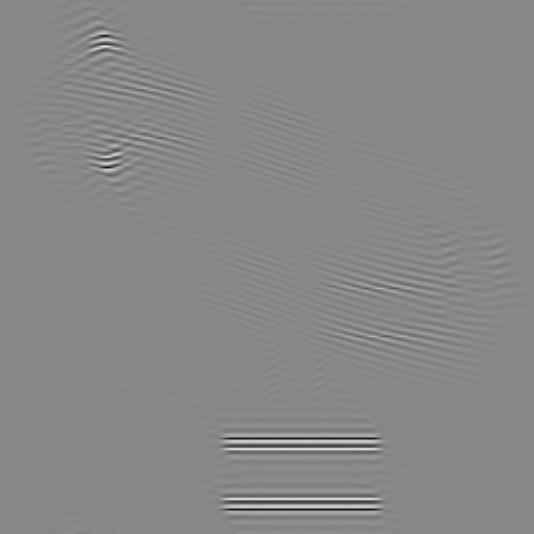} \\
\end{tabular}
\caption{Example of an empirical Voronoi wavelet transform. The top left image is an synthetically generated input image, in particular it contains four harmonic modes. The top right image depicted the detected Voronoi partition superimposed on the logarithm of the magnitude spectrum of the input image. The remaining images correspond to the outputs of the different empirical Voronoi wavelet filters. Note that the images that look black do actually contain some information of very small energy compared to the main harmonic modes.}
\label{fig:evwt}
\end{figure}

The example of an empirical Voronoi wavelet transform is given in Figure~\ref{fig:evwt}. The input image (on the top left of the figure) is a toy example made of piecewise objects (the oval and rectangle) on which four harmonic modes are superimposed. Two pairs of harmonic modes have similar frequencies but different orientations, i.e their corresponding positions in the Fourier domain lie in specific rings with different angular positions. The top right image shows the Voronoi partition plotted on top of the logarithm of the image magnitude spectrum. We can observe that the method indeed associates some specific cells to the particular harmonic modes. Finally, the wavelet coefficients, $f_k$, are given in the remaining images. We emphasize that each image has been re-normalized for visualization purposes most of them contain only information of very small magnitude compared to the main modes (given by the boxed images). On the other hand, the boxed images clearly show that some filters are indeed capable of extracting the different harmonic modes as well as the objects.

\section{Conclusion}
\label{sec:conc}
In this paper, we have proposed an alternative on how to create partitions in the Fourier domain for the purpose of building 2D empirical wavelets. Our solution, using Voronoi diagrams, provides a trade-off between having sub-domain with regular edges, and a high level of adaptability like the one previously proposed in the construction of empirical watershed wavelets. The corresponding Matlab code is publicly available at \url{https://www.mathworks.com/matlabcentral/fileexchange/42141-empirical-wavelet-transforms}.

\section{Acknowledgement}
This work was supported by the Air Force Office of Scientific Research under the grant number FA9550-21-1-0275.


\begin{thebibliography}{99}
			
		\bibitem{ewwt} {B. \ Hurat, Z. \ Alvarado {\rm and} J. \ Gilles}:
    	\textit{{The Empirical Watershed Wavelet}},
    	Journal of Imaging, {\bf 6} (12) (2020), 140.
%    	doi     = {10.3390/jimaging6120140}
       
        
		\bibitem{cewt} {J. \ Gilles}:
    	\textit{{Continuous empirical wavelets systems}},
		Advances in Data Science and Adaptive Analysis, {\bf 12} (03n04) (2020), 2050006.
%    	doi     = {10.1142/S2424922X20500060}

		\bibitem{microscopy} {K. \ Bui, J. \ Fauman, D. \ Kes, L.Torres \ Mandiola, A. \ Ciomaga, R. \ Salazar, A.L. \ Bertozzi, J. \ Gilles, D.P. \ Goronzy, A.I. \ Guttentag {\rm and} P.S. \ Weiss}:
    	\textit{{Segmentation of Scanning Tunneling Microscopy Images Using Variational Methods and Empirical Wavelets}},
    	Pattern Analysis and Applications, {\bf 23} (2020), 625--651.
%    	doi     = {10.1007/s10044-019-00824-0}
        
		\bibitem{supewttexture} {Y. \ Huang, F. \ Zhou {\rm and} J. \ Gilles}:
    	textit{{Empirical curvelet based Fully Convolutional Network for supervised texture image segmentation}},
		Neurocomputing, {\bf 349} (2019), 31--43.
%    	doi     = {10.1016/j.neucom.2019.04.021}
       
		\bibitem{ewttexture} {Y. \ Huang, V. \ De Bortoli, F. \ Zhou {\rm and} J. \ Gilles}:
    	textit{{Review of wavelet-based unsupervised texture segmentation, advantage of adaptive wavelets}},
    	IET Image Processing Journal, {\bf 12} (9) (2018), 1626--1638.
%    	doi     = {10.1049/iet-ipr.2017.1005}
        
		\bibitem{SSHS} {J.Gilles {\rm and} K. \ Heal}:
  		textit{A parameterless scale-space approach to find meaningful modes in histograms - Application to image and spectrum segmentation},
		International Journal of Wavelets, Multiresolution and Information Processing, {\bf 12} (6) (2014), 1450044-1--1450044-17.
%  		doi = {10.1142/S0219691314500441}
        
		\bibitem{EWT2D} {J. \ Gilles, G. \ Tran {\rm and} S. \ Osher}:
		textit{2D Empirical transforms. Wavelets, Ridgelets and Curvelets Revisited},
		SIAM Journal on Imaging Sciences, {\bf 7} (1) (2014), 157--186.
%  		doi     = {10.1137/130923774}
       
		\bibitem{EWT1D} {J. \ Gilles}:
    	textit{{Empirical Wavelet Transform}},
    	IEEE Transactions on Signal Processing, {\bf 61} (16) (2013), 3999--4010.
%    	doi     = {10.1109/TSP.2013.2265222}
        
		\bibitem{Huang1998} {N.E. \ Huang, Z. \ Shen, S.R. \ Long, M.C. \ Wu, H.H. \ Shih, Q. \ Zheng, N-C. \ Yen, C.C. \ Tung {\rm and} H.H. \ Liu}:
        textit{The empirical mode decomposition and the {Hilbert} spectrum for nonlinear and non-stationary time series analysis},
		Proc. Royal Society London A., {\bf 454} (1998), 903--995.
%        doi       = {10.1098/rspa.1998.0193}

		\bibitem{ewtapp1} {W. \ Liu, S. \ Cao {\rm and} Y. \ Chen}:
        textit{{Seismic Time–Frequency Analysis via Empirical Wavelet Transform}},
		IEEE Geoscience and Remote Sensing Letters, {\bf 13} (1) (2016),
        28--32.
%        doi     = {10.1109/LGRS.2015.2493198}

		\bibitem{ewtapp3} {X. \ Zhang, X. \ Li {\rm and} Y. \ Feng}:
        textit{{Image fusion based on simultaneous empirical wavelet transform}},
		Multimedia Tools and Applications, {\bf 76} (2017), 8175--8193.
%        doi     = {10.1007/s11042-016-3453-8}

		\bibitem{Otsu1979} {N. \ Otsu}:
        textit{A threshold selection method from gray-level histograms},
		IEEE Trans on Systems, Man and Cybernetics, {\bf 9} (1) (1979),
        62--66.
%        doi     = {10.1109/TSMC.1979.4310076}

		\bibitem{watershed1} {S. \ Beucher {\rm and} C. \ Lantu\'ejoul}:
        textit{Use of Watersheds in Contour Detection},
        International Workshop on Image Processing: Real-time edge and motion detection-estimation, Rennes, France, 2.1--2.12.

		\bibitem{watershed2} {F. \ Meyer}:
        textit{Topographic distance and watershed lines},
        Signal Processing, {\bf 38} (1) (1994), 113-125.
%        doi     = {10.1016/0165-1684(94)90060-4}

		\bibitem{watershed3} {F. \ Meyer {\rm and} S. \ Beucher}:
		textit{Morphological segmentation},
        Journal of Visual Communication and Image Representation, {\bf 1} (1) (1990), 21--46.
%        doi     = {10.1016/1047-3203(90)90014-M}

		\bibitem{voro1} {F. \ Aurenhammer, R. \ Klein {\rm and} D.-T. \ Lee}:
        textit{Voronoi Diagrams and Delaunay Triangulations},
        World Scientific, (2013).
%        doi     = {10.1142/8685}
\end{thebibliography}
\end{document}